%Tex file called: hc10.tex
%Contents: Preprint, Jan. 25, 1996.
%version: letter.
%
\def\gsim{\mathrel{\scriptstyle{\buildrel > \over \sim}}}

%\magnification=1095
\magnification 1200
\baselineskip=17pt
%\font\twelverm=cmr10 scaled 1200

%\vskip 32pt

%\centerline{\bf ENTROPIC CORRECTIONS TO THE FLUX-LINE ENERGY}

\centerline{\bf CRITICAL BEHAVIOR OF THE FLUX-LINE TENSION}
\bigskip
\centerline{\bf IN EXTREME TYPE-II SUPERCONDUCTORS}
\vskip 50pt
\centerline{J. P. Rodriguez}
\medskip
\centerline{Theoretical Division,
Los Alamos National Laboratory,
Los Alamos, NM 87545.\footnote*
{Permanent address: Dept. of Physics and Astronomy,
California State University,
Los Angeles, CA 90032.}}
\vskip 30pt
\centerline  {\bf  Abstract}
\vskip 8pt\noindent
The entropic corrections to the flux-line energy of extreme
type-II superconductors are computed using a schematic
dual Villain model description of the flux quanta.
We find that the temperature profile of the lower-critical
field vanishes polynomially at the transition with an exponent
$\nu\cong 2/3$ in the isotropic case, while it exhibits an
inflection point for the case of weakly coupled layers in 
parallel magnetic field.   It is argued that
vestiges of these effects have already been observed in 
high-temperature superconductors.
\bigskip
\noindent
PACS Indices: 74.20.De, 74.60.-w, 11.15.Ha, 75.10.Hk  
\vfill\eject

It is well known that type-II superconductors allow magnetic
fields to penetrate in the form of flux lines quantized in
units of the flux quantum, $\Phi_0 = hc/2e$, 
for field values above the lower
critical field given by $H_{c1}(T) = 4\pi\varepsilon(T)/\Phi_0$,
where $\varepsilon(T)$ denotes the flux-line free energy per
unit length.$^1$  The London approximation yields
$\varepsilon_0 (T) = (\Phi_0/4\pi\lambda_L)^2 {\rm ln}\, \lambda_L/\xi$ for
this free energy, where both the London penetration length, $\lambda_L$,
and the coherence length, $\xi$, vary as $(T_{c0}-T)^{-1/2}$
near the critical temperature, $T_{c0}$, within the Ginzburg-Landau
mean-field approximation.  Spatial fluctuations of
a flux line act to {\it reduce} this mean-field line tension, however,
particularly near the transition temperature.  Such entropic 
corrections to the lower critical field have been computed$^{2,3}$
and observed$^{4-6}$ in the case of high-temperature superconductors,
where fluctuation effects are important due to the short coherence
lengths and the large transition temperatures that
characterize these materials. 

In this paper, we shall re-compute the flux-line tension of extreme
type-II superconductors ($\lambda_L\gg\xi$)  in the absence of magnetic
field utilizing the so-called ``frozen'' 
limit of lattice superconductors first studied by Peskin.$^{7-9}$
The principle idea here is to exploit the duality of quantized
magnetic flux lines in superconductors to confining strings
of electric flux  in pure lattice gauge theories.
Hence, only the entropic correction
to the line tension due to flux-line wandering is accounted for.
The energy functional for three-dimensional (3D) superconductors
may be expressed as$^7$ 
$$E = \rho_s\sum_{r,\mu}\{1-{\rm cos}[\Delta_{\mu}\phi(r)-A_{\mu}(r)]\}
+{1\over{4 e_0^2}}\sum_{r,\mu,\nu} [\Delta_{\mu} A_{\nu}(r)
- \Delta_{\nu} A_{\mu}(r)]^2, \eqno (1)$$
where $\phi(r)$ is the phase of the order-parameter on a cubic
lattice with spacing $a^{\prime} > \xi$, and where
the magnetic flux threading the 
plaquette at site $r$  perpendicular to the
$\mu = x,y,z$ direction reads 
$\Phi_{\mu}(r) = (\Phi_0/2\pi) \sum_{\nu,\gamma} \epsilon_{\mu\nu\gamma}
\Delta_{\nu} A_{\gamma}$. 
Also, $\Delta_{\mu}\phi(r)=\phi(r+\hat\mu)-\phi(r)$
is the lattice difference operator.
A ``frozen'' superconductor (FZS) is simply the thermodynamic state in the
limit of infinite phase rigidity, $\rho_s\rightarrow\infty$.
This means that the London penetration length is small in
comparison to the effective lattice constant,  $a^{\prime}$,
and hence that
flux lines do not interact.  In particular,
upon making the gauge transformation $A_{\mu} = A_{\mu}^{\prime}
+ \Delta_{\mu}\phi$, we immediately see that this limit
implies $A_{\mu}^{\prime} = 2\pi m_{\mu}$, where $m_{\mu}(r)$
is an integer field.  Therefore, the energy functional (1) reduces to
$E_{\rm FZ} = \sum_{r,\mu} (2 e_0)^{-2} [2\pi n_{\mu}(r)]^2$, 
where $n_{\mu}(r)$
is the integer field describing unit line segments of quantized
flux traversing the  plaquette $(r,\mu)$
and satisfying the continuity equation
$$\sum_{\mu} \Delta_{\mu} n_{\mu}|_r = 0; \eqno (2)$$
i.e., $n_{\mu} = \sum_{\nu,\gamma} \epsilon_{\mu\nu\gamma}
\Delta_{\nu} m_{\gamma}$.		
The universality class of the transition into the normal state 
thus falls within that of the {\it inverted} 3D $XY$ model.$^{7-9}$
Employing this schematic description of flux lines in the absence of
external magnetic field, with an effective lattice constant 
$a^{\prime} \gsim  \lambda_L$,
we shall demonstrate below that the lower critical
field vanishes as $H_{c1}\propto (T_c-T)^{\nu}$ near the transition
in ``frozen'' lattice superconductors,
where the exponent $\nu\cong 2/3$ corresponds to that of the 3D
$XY$ model.  Via an anisotropic version of this FZS model,$^{10}$
we shall also show that the temperature
profile for the parallel lower critical field 
in superconducting films with weakly coupled layers
has an inflection point. 
Last, we argue that these effects are realized in
high-temperature superconductors in the critical regime.$^{4-6}$ 

%near $T_{c0}$ and at low magnetic fields

{\it Isotropic Superconductor.}  Let us then describe
the zero-field thermodynamics of flux lines in 3D superconductors
via the energy functional
$$E_{\rm FZ}/k_B T = \beta \sum_{r,\mu} n_{\mu}^2(r) \eqno (3)$$
along with constraint (2), where $n_{\mu}(r)$ is the integer field
that represents a unit line segment of magnetic flux quanta threading
the plaquette $(r,\mu)$.  The energy scale above is set by the
mean-field flux-line free energy, 
$\varepsilon_0(T) = (\Phi_0/4\pi\lambda_0)^2 (1-T/T_{c0})$,
and the effective lattice constant, $a^{\prime}$, yielding 
$\beta = a^{\prime} \varepsilon_0 (T)/ k_B T$.  Here,
$\lambda_0$ is on the order of the zero-temperature London
penetration length, while $T_{c0}$ denotes the mean-field critical
temperature.  We presume that the lattice constant satisfies
$a^{\prime}\gsim \lambda_L$, which insures that the magnetic flux
is quantized on the scale of an elementary plaquette.
To compute the dimensionless flux-line tension,
$\sigma$, we shall place a unit monopole/anti-monopole pair separated
by a large distance $R\gg a^{\prime}$ along a principle axis of
the cubic lattice.  The partition function for the FZS system (3)
in the presence of such an external monopole charge configuration,
$p(r) = \delta_{r,0} - \delta_{r, R\hat x}$, therefore reads
$$Z[p] = \sum_{\{n_{\mu}(r)\}}\Pi_{r}
\delta\Bigl[\sum_{\mu} \Delta_{\mu} n_{\mu}|_{r} - p(r)\Bigr]
{\rm exp}(-E_{\rm FZ}/k_B T). \eqno (4)$$
Yet since $Z[p]/Z[0] \propto e^{-\beta\sigma R}$ 
coincides with the correlation function of the
dual 3D $XY$ model in the Villain form,$^7$
$C(R) \propto e^{-R/\bar\xi}$, the 
dimensionless flux-line
free energy is then related to 
the correlation length, $\bar\xi$, of the latter
by $\sigma = (\beta\bar\xi)^{-1}$.  
(Henceforth, $R$ and $\bar\xi$ are given in units of $a^{\prime}$.)
This implies that
the line tension varies as $\sigma = \sigma_0(\beta - \beta_c)^{\nu}$
near the critical temperature, where $\nu\cong 2/3$ is the critical
exponent corresponding to the divergent correlation length of the
3D Villain model.$^7$  Here, $\beta_c\cong {3\over 2}$ is the critical
temperature of the latter,$^9$ while $\sigma_0$ is another non-universal
constant of order unity.
Since the dimensions of a typical flux loop are set by the
divergent correlation length, $\bar\xi$, short-range interactions among
the flux lines on the scale of the {\it finite}   mean-field penetration
length should be irrelevant at criticality.  This validates
the FZS limit (3).
The flux-line tension (or $Z[p]/Z[0]$) can also be
computed at low temperature by permitting maximum transverse
excursions in  the line of one lattice spacing in any of the
remaining $2(d-1)$ directions,  
where $d=3$ gives the dimension of the cubic  lattice.
This  yields  
$$\sigma\cong 1 - 2(d-1)\beta^{-1} e^{-\beta}, \eqno (5)$$
which agrees with the low-temperature dependence of
the interface tension in the  two-dimensional (2D) 
Ising model for the case $d=2$,$^{11}$ as it should.

The mean-field flux-line free energy in the ``frozen'' superconductor model
[(2) and (3)] is therefore renormalized {\it down} by spatial
fluctuations to $\varepsilon (T) = \varepsilon_0(T) \sigma(\beta)\propto
(\beta-\beta_c)^{\nu}$ near the critical temperature, 
$T_c = (T_{c0}^{-1}+\beta_c T_0^{-1})^{-1}$, which
is  determined by the condition
$\beta = \beta_c$.  Here, the energy scale
$k_B T_0 = a^{\prime} (\Phi_0/4\pi\lambda_0)^2$
sets the size of the critical regime,
$\delta T_c = T_{c0}-T_c\sim T_{c0}^2/T_0$.  
In Fig. 1, we plot the predicted 
profile of the lower critical field vs. temperature,
which is determined by the 
flux-line tension in zero-field via $H_{c1} = 4\pi\varepsilon/\Phi_0$
in the case of extreme type-II superconductors.
Here, we have interpolated between the low-temperature and 
critical behaviors discussed previously.
Last, we mention that a $\lambda$-transition is
also expected in the specific heat vs. temperature.$^{7,9}$
 
{\it Layered Superconductor (Film).}  Consider now an anisotropic FZS
model
made up of $N$ weakly coupled
square lattices separated by the same 
effective lattice constant $a^{\prime} \gsim \lambda_0$,
with an energy functional given by
$$E_{\rm FZ}/k_B T = \beta\sum_{l=1}^N\sum_{\vec r}|\vec n (\vec r, l)|^2
+\gamma\beta\sum_{l=1}^{N-1}\sum_{\vec r} |n_z(\vec r, l)|^2 \eqno (6)$$
along with constraint (2). 
Here, $\vec n = (n_x, n_y)$, $\vec r$ and $l$ respectively denote the
in-plane and layer indices, while   $\gamma\gg 1$
denotes the anisotropy parameter.$^1$  The smallest energy scale
is set by the mean-field in-plane flux-line tension,$^1$ 
$\gamma^{-1}\varepsilon_0(T)$, and the lattice constant, 
yielding $\beta = \gamma^{-1}\varepsilon_0(T) a^{\prime}/k_B T$.
The author has recently studied this model in the context of
the dual $XY$ model in the Villain form with a finite number
of weakly coupled layers,$^{10}$ where he finds 
a superfluid ordering transition at (inverted) temperature
$\beta_c\sim\pi/4$, followed by a layer-decoupling transition
at higher (inverted) temperature $\beta_* = 2\pi$.
Very similar results were obtained previously by Korshunov
in his approximate study of the infinite-layer case.$^{12,13}$
Specifically, the in-plane field may be expressed as
$\vec n(\vec r, l) = \vec n\,^{\prime}(\vec r, l)
- \vec n_-(\vec r, l) + \vec n_-(\vec r, l-1)$,
along with constraints
$\vec\nabla\cdot\vec n^{\prime} = p$ and
$\vec\nabla\cdot\vec n_- = n_z$, where
$p(r)$ again represents the monopole distribution and
$\vec\nabla = (\Delta_x, \Delta_y)$.
Employing the potential representation $\vec n_- = - \vec\nabla \Phi$
factorizes the partition function (4) into 
$Z = Z_{\rm CG}\Pi_{l=1}^{N} Z_{\rm DG}^{(l)}$ in the limit
$\gamma\rightarrow\infty$,$^{10}$ where $Z_{\rm DG}^{(l)}$
is the partition function for  the 2D discrete gaussian model$^{14}$
that describes the $l^{\rm th}$ layer isolated from its neighbors
($n_z = 0$), while the inter-layer Coulomb gas factor reads
$$\eqalignno{
Z_{\rm CG} [p] =  \sum_{\{ n_z(\vec r, l)\}} {\rm exp}
\Biggl\{-\beta\sum_{l=1}^{N}
&\sum_{\vec r, \vec r\,^{\prime}}[n_z(\vec r, l-1) - n_z(\vec r, l) 
+ 2 p(\vec r, l)]
G^{(2)}(\vec r - \vec r \,^{\prime})\times \cr
&\times [n_z(\vec r\,^{\prime}, l-1) - n_z(\vec r\,^{\prime}, l)]
-\gamma\beta\sum_{l=1}^{N-1}
\sum_{\vec r} n_z^2(\vec r, l)\Biggr\},& (7) \cr}$$
with the boundary layers set to
$n_z(\vec r, 0) = 0 = n_z(\vec r, N)$.
Here, $G^{(2)} = -\nabla^{-2}$
% (\vec r)= (2\pi)^{-2}
% \int_{\rm BZ}d^2k (e^{i\vec k\cdot \vec r}-1)
% [4-2 {\rm cos} (k_x  a^{\prime}) -2 {\rm cos} (k_y a^{\prime})]^{-1}$ 
is the Greens function
for the square lattice.  Notice that flux lines passing vertically through
all of the layers make no contribution to the energy functional above.$^{15}$
In the limit of extreme anisotropy $\gamma\rightarrow\infty$, we thus find
that $Z_{\rm DG}^{(l)}$ and $Z_{\rm CG}$ respectively exhibit an inverted 
Kosterlitz-Thouless (KT) transition at $\beta = \beta_c$ and
a Coulomb gas$^{16}$ transition at $\beta = \beta_*\sim 10\beta_c$.

To compute the parallel flux-line tension, we again place unit
oppositely charged monopoles separated by a large distance $R\gg a^{\prime}$
along a principle axis of a given layer $l$.  Since the
corresponding intra-layer
correlation function of the dual Villain model factorizes into
$C(R) = C_{\parallel}(R) C_{\perp}(R)$, where
$C_{\parallel}(R) = Z_{\rm DG}^{(l)}[p]/Z_{\rm DG}^{(l)}[0]
= (r_0/R)^{\eta_{\parallel}}{\rm exp}(-R/\bar\xi_{\parallel})$ and
$C_{\perp}(R) = Z_{\rm CG}[p]/Z_{\rm CG}[0]
= (R/r_0)^{\eta_{\perp}}$, we obtain that the dimensionless
parallel line tension is given by $\sigma = (\beta\bar\xi_{\parallel})^{-1}$,
where $\bar\xi_{\parallel}$ denotes the correlation length of
the 2D Villain model$^{16}$ dual to the former discrete-gaussian
model.$^{10,14}$  Hence, this line tension vanishes as 
$\sigma = \sigma_0\,{\rm exp}[-A_0/(\beta-\beta_c)^{1/2}]$ just
below the critical temperature,$^{14}$
$T_c = (T_{c0}^{-1}+\beta_c\gamma T_0^{-1})^{-1}$,
where $\sigma_0$ and $A_0$ again are non-universal
constants of order unity.  
A similar temperature dependence for the parallel lower-critical field
in layered superconductors
has been obtained via a fermion analogy for the Lawrence-Doniach
model.$^{13}$
The low-temperature behavior of the parallel line tension,
on the other hand,  is
determined by Eq. (5) for $d=2$.  
Although the corresponding
parallel flux-line free energy, $\varepsilon_{\parallel} (T) =
\gamma^{-1}\varepsilon_0 (T)\sigma(\beta)$, is renormalized
down with respect to the mean-field result just like in the
isotropic case, we also observe that it must have an inflection
point as a function of temperature just below the transition,
where it vanishes exponentially.     This is displayed by
the parallel lower-critical field ($H_{c1}^{\parallel}$)
in Fig. 2,  which interpolates
between the low-temperature and critical behaviors. 
The exponential suppression of $H_{c1}^{\parallel}$ near $T_c$
indicates that critical  flux-line wandering is 
extremely violent in two dimensions.

%which is again set by the zero-field line tension via 
%$H_{c1}^{\parallel} = 4\pi\varepsilon_{\parallel}/\Phi_0$. 
%in extreme type-II superconductors. 

Concerning the lower-critical field $H_{c1}^{\perp}$
perpendicular to the layers, classical electrodynamics 
dictates that it must vanish in the present film
geometry.  However, we may fix the ends of a vertical flux line
at the boundary layers 
[$n_z(\vec r,0) = \delta_{\vec r, 0} = n_z(\vec r, N) $] and compute
its mean-squared displacement
$\bar u_l^2 = \sum_{\vec r\neq 0} r^2\langle  n_z(\vec r, l) n_z(0, l)\rangle
/\sum_{\vec r\neq 0} \langle  n_z(\vec r, l) n_z(0, l)\rangle$ 
at a given layer $l$.  
In the limit of extreme
anisotropy, $\gamma\rightarrow\infty$, it becomes sufficient
to consider the double-layer case, $N=2$. 
Since the charge correlation function for the 2D Coulomb gas (7)
satisfies$^{16}$ 
$\langle n_z(0) n_z(\vec r)\rangle\propto - r^{-2\beta/\pi}$,
we obtain
$\bar u^2 = a^{\prime 2}(\beta - {1\over 2}\beta_*)/(\beta - \beta_*)$.
Exactly the same result is also obtained
for pancake vortices in the 
absence of Josephson coupling between layers$^{3,17}$ 
if the in-plane London penetration length  $\lambda_L$ is chosen for the 
lattice constant $a^{\prime}$.  The latter choice insures
flux quantization at all temperatures. It also indicates that
an elementary segment of vertical flux line, $n_z = \pm 1$,
corresponds to a {\it stack} of pancakes.
We may therefore interpret the lower transition temperature, 
$T_* = (T_{c0}^{-1}+\beta_*\gamma T_0^{-1})^{-1}$,
determined by $\beta = \beta_* = 2\pi$
%\sim 10\beta_c$ 
as an evaporation
point for vertical flux lines (see Fig. 2), since $\bar u^2$
diverges for $T>T_*$.
If $\gamma$ is now presumed to be large but finite, however,
we must then require that the perpendicular magnetic field
exceed the decoupling  cross-over field at $T_*$,$^{18}$
$B_*^{\perp}\sim\Phi_0/\gamma^2 d^2$. This   guarantees that the
distance between neighboring  stacks of
pancake  vortices be within the Josephson 
penetration length,
$\gamma d$.  Here $d$ measures the distance between adjacent Josephson-coupled
planes in the underlying Lawrence-Doniach model for layered
superconductivity,$^1$ which is essentially an anisotropic version
of Eq. (1). 
The former condition excludes the possibility that a well-defined
Josephson vortex   connects displaced stacks of pancake vortices in adjacent
layers, an effect that is not
accounted for by the present factorization of the layered FZS model
(6) into parallel and perpendicular parts.   
Last, we note that compatibility with the present FZS model requires
that the physical length scales satisfy
$\gamma d\gg a^{\prime}\gsim\lambda_0$.

The above  results for the  FZS
model (6) are in apparent contradiction with those 
of the more general anisotropic lattice superconductor model (1)
obtained by  Korshunov,$^{12}$
who finds that layers decouple at a temperature
tenfold {\it greater} than the superfluid KT transition.
This paradox is resolved by observing ({\it i}) that the
lattice constant in such models is on the order of the
coherence length, $\xi\ll\lambda_L$, and ({\it ii}) that such decoupling
of layers naturally coincides with the vanishing of
the parallel lower-critical field here.  Notice 
this means that in the regime $T_* < T < T_c$,
a Meissner phase exists for in-plane 
magnetic fields while each layer is {\it resistive},
a feature which has been observed experimentally.$^{6,13}$  
The latter, however, is then consistent
with the ``evaporation'' of vertical flux lines at $T_*$
obtained here.  Last, we mention that the 
line tension for parallel vortices
can be computed directly from the double-layer
$XY$ model,$^{10}$  where one recovers the previous
exponential temperature dependence shown in Fig. 2.$^{19}$ 
Again, this demonstrates the consistency of the present
results with those obtained from model (1) in the extreme
type-II limit, $\lambda_L\rightarrow\infty$, which is 
precisely the $XY$-model.

{\it High-Temperature Superconductors.}  Let us now apply the
above theory to the oxide superconductors,
which are extreme type-II ($\lambda_L/\xi\sim 100$)
with observable fluctuation effects.$^{1-6}$  Consider first
bulk YBa$_2$Cu$_3$O$_{7-y}$, which may be classified as a 3D
superconductor with strong anisotropy,$^1$ $\gamma\sim 10$.
The {\it scaled} 
isotropic FZS model (3) predicts a width in temperature for
critical fluctuations (see Fig. 1) on the order of 
$\delta T_c\sim \gamma T_{c0}^2/T_0$, 
where again $k_B T_0=(\Phi_0/4\pi\lambda_0)^2 a^{\prime}$.
Assuming typical parameters $T_{c0}\sim 100\,{\rm K}$
and $a^{\prime}\sim\lambda_0\sim 10^3\,{\rm \AA}$, we obtain
$\delta T_c\sim 0.5\,{\rm K}$.  This estimate is consistent
with recent $H_{c1}$ measurements performed  by Pastoriza et al.,$^4$
who observe  fluctuation effects {\it \` a la} Fig.  1
to within this scale.  
If we suppose, however, that
$a^{\prime}\sim\lambda_L(T) = \lambda_0 (T_{c0}/\delta T_c)^{1/2}$,
which insures flux quantization at all temperatures,
then we obtain a much more modest estimate    of
$\delta T_c\sim 2.5\,{\rm mK}$ for the size of the
critical region.
Also, the exponent $\nu=2/3$ predicted here for the
critical temperature dependence of the
lower-critical field is consistent with the critical temperature
dependence of the London penetration length  $\lambda_L$ per
the (neutral) 3D $XY$ model;$^1$
i.e., $H_{c1}\propto \lambda_L^{-2}$, with
$\lambda_L\propto (T_c-T)^{-1/3}$.
The latter critical temperature dependence has been 
observed in recent  penetration depth measurements on
this class of high-temperature
superconductors.$^{20}$

The oxide superconductor
Bi$_2$Sr$_2$CaCu$_2$O$_8$, on the other hand, 
may be classified
as a layered superconductor with extreme anisotropy,$^1$
$\gamma\sim 100$.  The layered FZS model (6) also predicts
a fluctuation regime on the order of
$\delta T_c\sim \gamma T_{c0}^2/T_0$ wide in temperature.
Taking the previous values for $\lambda_0$
and $T_{c0}$, then the choice $a^{\prime}\sim \lambda_0$
results in $\delta T_c\sim 5\,{\rm K}$, which 
is on the order of the effect (see Figs. 1 and 2)
observed by Brawner et al.$^5$ and
by Wan et al.$^6$ in their measurements of the field of first
penetration, $H_p$, along the (perpendicular) c-axis for this material.  
On the other hand, the choice 
$a^{\prime}\sim \lambda_L(T) = \lambda_0 (T_{c0}/\delta T_c)^{1/2}$ for
the lattice constant with the same parameter values
results in a smaller estimate
$\delta T_c\sim 0.25\,{\rm K}$ for the size of the
critical region.
More interestingly,
Wan et al. also observe an inflection point in $H_p$ vs. $T$ on this
scale. This  may be a vestige of the same prediction made here for
the parallel lower-critical field if
geometrical  demagnetization effects
are presumed to be strong.  In particular, only parallel flux lines
contribute to the total line 
free energy at temperatures $T_* < T < T_c$,
since the perpendicular flux lines
have by then ``evaporated''.
This scenario requires, of course, that the perpendicular magnetic field
exceed the decoupling cross-over scale,$^{18}$
$B_*^{\perp}\sim\Phi_0/\lambda_J^2$, where $\lambda_J$ represents
the Josephson penetration length.  Since this scale is known to
diverge exponentially as $T$ approaches the decoupling
transition temperature $T_c$ from below,$^{19}$ the former cross-over
scale should be negligibly small in that regime.  At the evaporation
point ($T=T_*$), however, we have  that $\lambda_J\sim \gamma d$. 
This  implies a
decoupling cross-over field of    
approximately $1\,{\rm kG}$ in the present case,$^{1}$ where
$d = 15\,{\rm\AA}$.
A direct measurement of the parallel lower-critical field in such
material would clearly be very useful.  Last, we mention that
Wan et al. have fit their results for the 
2D-3D crossover line of c-axis aligned flux lines 
to$^6$ $H_m\propto (\beta-\beta_c)^{\nu}$, with $\nu = 0.68$.
The latter exponent is precisely that predicted for $H_{c1}$ vs. $T$
in the isotropic FZS model (3), which suggests that
$H_m$ is related in some way to {\it bulk}
$H_{c1}$ along the c-axis.

Finally, we note that the Coulomb gas model (7) for vertical
flux-line segments indicates that the evaporation transition
at $\beta^{-1} = \beta_*^{-1}\sim \beta_c^{-1}/10$ occurring in the limit
$\gamma\rightarrow\infty$ is renormalized down to
$\beta^{-1} = \epsilon_0^{-(N-1)} \beta_*^{-1}$ for $\gamma$ finite,
where $\epsilon_0 - 1\propto e^{-2\gamma\beta_*}$  is the polarization
at the transition point 
due to flux-line/anti-flux-line excitations 
in between consecutive layers. 
Hence, flux line ``evaporation'' in clean layered superconductors
is driven to lower temperature as the c-axis thickness increases.$^{21}$
If we presume the existence of strong columnar pins, however,
then such isolated flux-line/anti-flux-line excitations between
layers are suppressed, resulting in an unrenormalized flux-line
evaporation transition.  Precisely this effect has been observed
in high-temperature superconductors, where the 
melting-evaporation-irreversibility line moves up in temperature
upon the introduction of columnar pins.$^{22}$

In summary, we have elucidated the critical behavior of the
flux-line tension in both isotropic and layered extreme
type-II superconductors ($\xi\rightarrow 0$) 
utilizing the so-called ``frozen''
limit of a lattice model.$^{7-9}$  Notably, we predict an
inflection point near the transition for the parallel lower-critical
field as a function of temperature in layered superconducting films.
Vestiges of this effect have already appeared in quasi-2D oxide
superconductors.$^6$  It is important to remark that fluctuations
in the magnitude of the superconducting order-parameter are 
entirely neglected within this approach, contrary to a recent
treatment of the same problem.$^{23}$   This is generally
a valid approximation 
when effects due to the normal cores are  negligible,
like in the present limit $\xi\rightarrow 0$, and
in quasi-2D systems, where phase fluctuations
drive critical properties.  

%We close by mentioning that
%measurements of the in-plane lower-critical field in films
%of oxide superconductors would be highly desirable.

The author is indebted to H. Safar, M. Maley,
S. Fleshler, G. Gilmer and  L. Bulaevskii
for valuable discussions.    This work
was performed under the auspices of the U.S. Department of Energy,
and  was supported in part by
National
Science Foundation grant DMR-9322427.

\vfill\eject
\centerline{\bf References}
\vskip 16 pt

\item {1.}  G. Blatter, M.V. Feigel'man, V.B. Geshkenbein, A.I. Larkin,
and V.M. Vinokur, Rev. Mod. Phys. {\rm 66}, 1125 (1994), 
and references therein.

\item {2.}  D.R. Nelson and H.S. Seung, Phys. Rev. B {\bf 39}, 9153 (1989).

\item {3.}  L.N. Bulaevskii, S.V. Meshkov, and D. Feinberg, 
Phys. Rev. B {\bf 43}, 3728 (1991);  L.N. Bulaevskii, M. Ledvij, and
V.G. Kogan, Phys. Rev. Lett. {\bf 68}, 3773 (1992).

\item {4.}  H. Safar, H. Pastoriza, F. de la Cruz, D.J. Bishop, 
L.F. Schneemeyer, and J.V. Waszczak, Phys. Rev. B {\bf 43}, 13610 (1991);
H. Pastoriza, H. Safar, E. Fern\' andez Righi, and F.
de la Cruz, Physica B {\bf 194}, 2237 (1994).

\item {5.}  D.A. Brawner, A. Schilling, H.R. Ott, R.J. Haug, K. Ploog,
and K. von Klitzing, Phys. Rev. Lett. {\bf 71}, 785 (1993).

\item {6.}  Y.M. Wan, S.E. Hebboul, and J.C. Garland, Phys. Rev. Lett.
{\bf 72}, 3867 (1994).

\item {7.}  M.E. Peskin, Ann. of Phys. {\bf 113}, 122 (1978).

\item {8.}  P.R. Thomas and M. Stone, Nucl. Phys. B {\bf 144},
513 (1978).

\item {9.}  C. Dasgupta and B.I. Halperin, Phys. Rev. Lett. {\bf 47},
1556 (1981).

\item {10.}  J.P. Rodriguez, Europhys.   Lett. {\bf 31}, 479 (1995).
% (cond-mat/9501033).

\item {11.}  L. Onsager, Phys. Rev. {\bf 65}, 117 (1944);
M.E. Fisher and A.E. Ferdinand, Phys. Rev. Lett. {\bf 19}, 169 (1967).

\item {12.}  S.E. Korshunov, Europhys. Lett. {\bf 11}, 757 (1990);
see also G. Carneiro, Physica C {\bf 183}, 360 (1991).

\item {13.}  B. Horovitz, Phys. Rev B {\bf 47}, 5947 (1993); 5964
(1993); Phys. Rev. Lett. {\bf 72}, 1569 (1994).

\item {14.} H. van Beijeren and I. Nolden, in
{\it Structure and Dynamics of Surfaces II},
edited by W. Schommers
and P. von Blanckenhagen (Springer, Heidelberg, 1987).

\item {15.}  A model for vortices in layered systems with a form
similar to that of Eq. (7) has been recently studied by
S.W. Pierson, Phys. Rev. Lett. {\bf 73}, 2496 (1994).

\item {16.} J.V. Jos\' e, L.P. Kadanoff, S. Kirkpatrick and
D.R. Nelson, Phys. Rev. B {\bf 16}, 1217 (1977).

\item {17.}  J.R. Clem, Phys. Rev. B {\bf 43}, 7837 (1991);
K.H. Fischer, Physica C {\bf 178}, 161 (1991).

\item {18.} L.I. Glazman and A.E. Koshelev, Phys. Rev. B {\bf 43},
2835 (1991).

\item {19.}  J.P. Rodriguez, Los Alamos report \# LA-UR-95-1908.

\item {20.} S. Kamal, D.A. Bonn, N. Goldenfeld, P.J. Hirschfeld, 
R. Liang, and W.N. Hardy,
Phys. Rev. Lett. {\bf 73}, 1845 (1994).

\item {21.}  Arguments based on duality with the layered $XY$ model
indicate that the evaporation transition is suppressed to low temperature
for $N\gsim\gamma^{1/2}$ (see ref. 10).

\item {22.}  D.R. Nelson and V.M. Vinokur, Phys. Rev. B {\bf 48},
13060 (1993), and references therein.

\item {23.} G. Blatter, B. Ivlev, and H. Nordborg, Phys. Rev. B {\bf 48},
10448 (1993).

\vfill\eject
\centerline{\bf Figure Captions}
\vskip 20pt
\item {Fig. 1.}   Shown is a temperature profile for $H_{c1}$  
(solid line) characteristic
of isotropic extreme type-II superconductors in the critical regime
(shaded area).
Notably, $H_{c1}$ deviates from the mean-field result (dashed line)
in an activated manner, and it approaches the transition temperature
by following a polynomial law with exponent $\nu\cong 2/3$
(see ref. 7). 

%The latter result is implicit in ref. 7. 

\item {Fig. 2.}  The parallel lower-critical field characteristic
of extreme type-II superconducting films that are composed of
weakly coupled layers
is plotted as a function of temperature near the phase transition.
Notice that it vanishes exponentially beyond
the inflection point just below $T_c$.
Notice also that the 
low-temperature transition at $T_*$, 
which marks where perpendicular flux lines 
``evaporate'', lies well inside the mean-field regime.
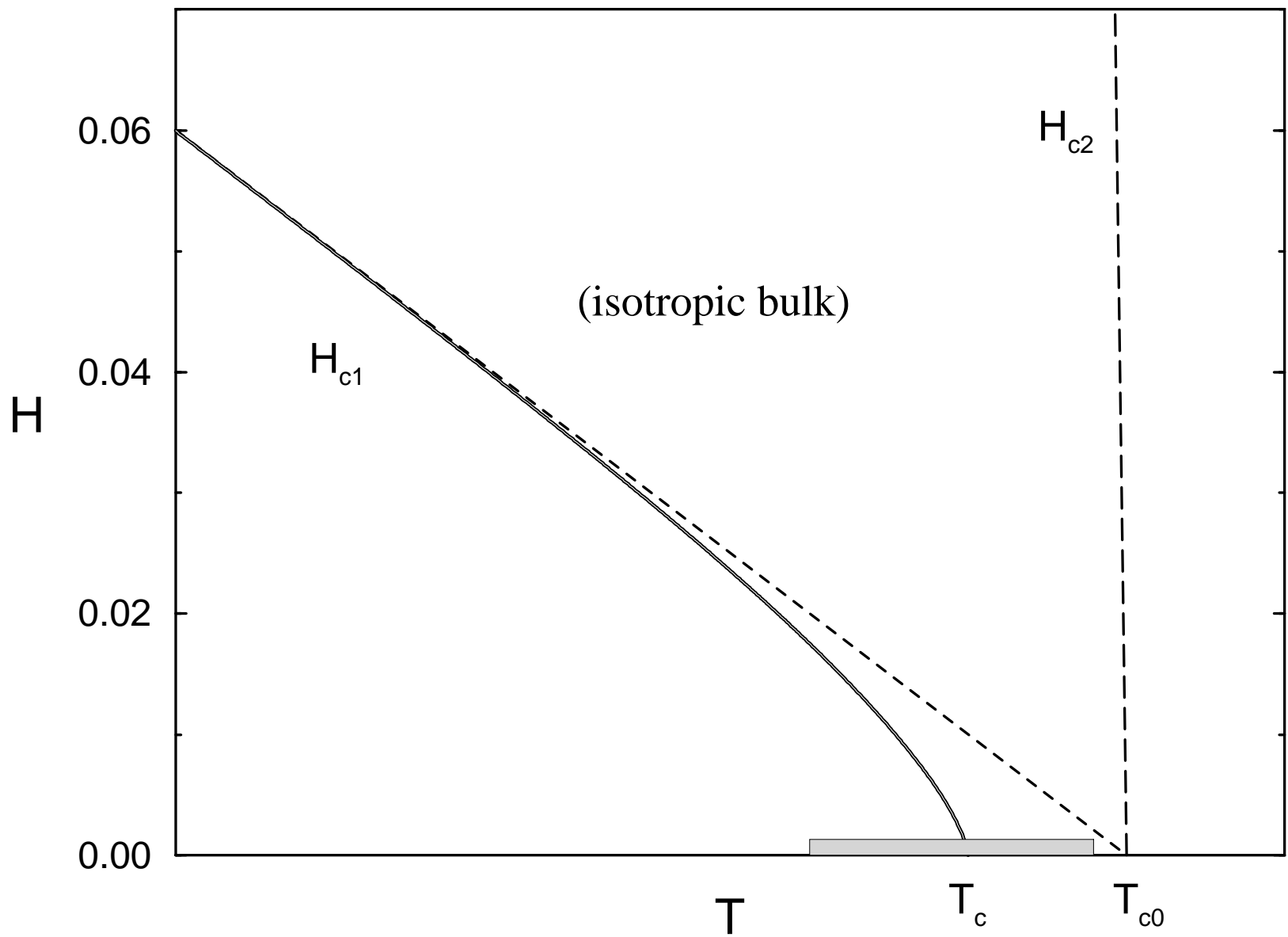
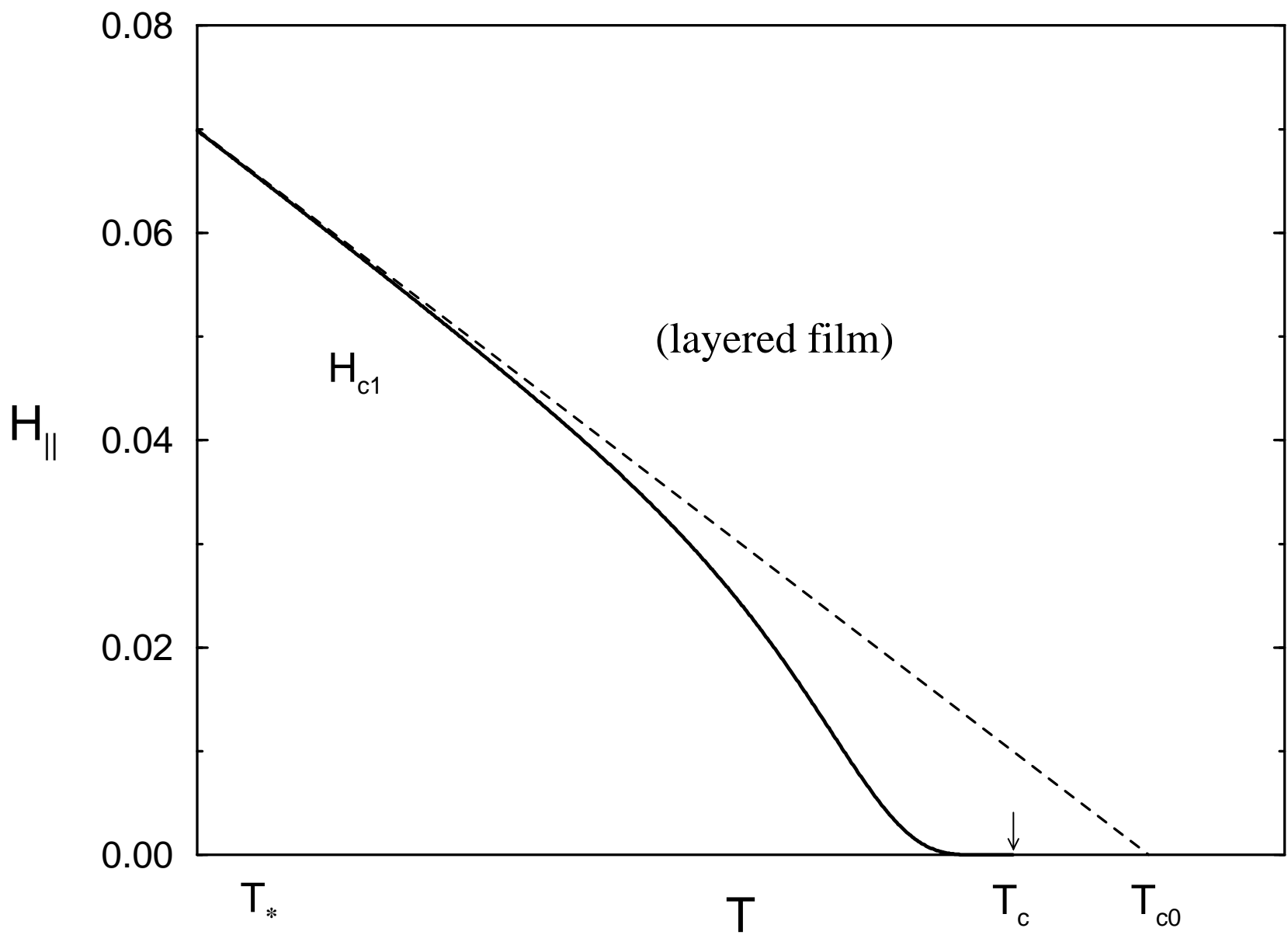
\end